\documentclass[fleqn, usenatbib]{mnras}
\usepackage{newtxtext,newtxmath}

\usepackage[T1]{fontenc}
\usepackage{ae,aecompl}

\usepackage{graphicx}	
\usepackage{amsmath}	
\usepackage{amssymb}	

\title[UV habitability of Proxima $b$] {The UV surface habitability of Proxima $b$: first experiments revealing probable life survival to stellar flares}
 
\author[X. C. Abrevaya et al.]{
X. C. Abrevaya,$^{1,2,3}$\thanks{E-mail: abrevaya@iafe.uba.ar}\thanks{Visiting scientist at the University of Graz}
M. Leitzinger,$^{3}$ 
O. J. Oppezzo,$^{4}$ 
P. Odert,$^{3}$
M. R. Patel,$^{5}$
\newauthor
G. J. M. Luna,$^{1,2}$
A. F. Forte-Giacobone,$^{4,6}$ 
and A. Hanslmeier $^{3}$
\\
$^{1}$Instituto de Astronom\'{\i}a y F\'{\i}sica del Espacio (UBA-CONICET) Ciudad Aut\'onoma de Buenos Aires, Argentina\\
$^{2}$Facultad de Ciencias Exactas y Naturales, Universidad de Buenos Aires Ciudad Aut\'onoma de Buenos Aires, Argentina\\
$^{3}$Institute of Physics, IGAM, University of Graz, Austria\\
$^{4}$Departamento de Radiobiolog\'{\i}a, Comisi\'on Nacional de Energ\'{\i}a At\'omica. Buenos Aires, Argentina\\
$^{5}$School of Physical Sciences, The Open University, Milton Keynes, MK7 6AA, U.K.\\
$^{6}$Departamento de Ingenier\'{\i}a, Universidad Nacional de Tres de Febrero. Buenos Aires, Argentina
}

\date{Accepted XXX. Received YYY; in original form ZZZ}

\pubyear{2019}

\begin{document}
\label{firstpage}
\pagerange{\pageref{firstpage}--\pageref{lastpage}}
\maketitle

\begin{abstract}
We use a new interdisciplinary approach to study the UV surface habitability of Proxima $b$ under quiescent and flaring stellar conditions. We assumed planetary atmospheric compositions based on CO$_2$ and N$_2$ and surface pressures from 100 to 5000 mbar. Our results show that the combination of these atmospheric compositions and pressures provide enough shielding from the most damaging UV wavelengths, expanding the ''UV-protective'' planetary atmospheric compositions beyond ozone. Additionally, we show that the UV radiation reaching the surface of Proxima $b$ during quiescent conditions would be negligible from the biological point of view, even without an atmosphere. Given that high UV fluxes could challenge the existence of life, then, we experimentally tested the effect that flares would have on microorganisms in a ''worst case scenario'' (no UV-shielding). Our results show the impact that a typical flare and a superflare would have on life: when microorganisms receive very high fluences of UVC, such as those expected to reach the surface of Proxima b after a typical flare or a superflare, a fraction of the population is able to survive. Our study suggests that life could cope with highly UV irradiated environments in exoplanets under conditions that cannot be found on Earth.
\end{abstract}

\begin{keywords}
Astrobiology -- 
   Planets and satellites: terrestrial planets -- 
   Planets and satellites: surfaces -- 
   Stars: flare -- 
   Stars: activity -- 
   Ultraviolet: stars
\end{keywords}



\section{Introduction}
Stellar ultraviolet radiation (UVR) is an important factor that can influence the existence of ''life as we know it'' on the surface of planetary bodies \citep[for review on the topic see:][]{Hanslmeier2018, Abrevaya2018}. Although several studies have focused on stellar UVR aiming to evaluate the chances of an exoplanet to host life on its surface, they are based on modelling \citep[see e.g.:][]{Cockell1999, Buccino2006, Buccino2007, Fossati2012, OMalley2017, Howard2018, Estrela2018, Rugheimer2015, OMalley2019}. Some of these approaches have neglected some important factors that may lead to: (i) an inaccurate estimation of the UV flux on the surface of the planet, as some studies do not assume the presence of a planetary atmosphere or assume unrealistic atmospheric compositions for the stellar spectral type and activity \citep[e.g.: O$_2$ rich Earth-like atmospheric compositions are unfeasible in planets orbiting M dwarfs, e.g.:][]{Johnstone2019} (ii) an inaccurate estimation of the biological impact of UV by using extrapolations of biologically relevant UV fluences\footnote{Defined according to \citet{Braslavsky2007} as: at a given point in space, the radiant energy incident on a small sphere from all directions divided by the cross-sectional area of that sphere. SI unit J~m$^{-2}$} based on empirical data from the literature, such as the UV biological action spectra (UV-BAS; a measure of biological UV damage at a given  UV  wavelength under a given constant fluence), or the UV ''lethal dose'' (UV-LD; the estimated amount of UVR required to kill a determined percentage of the population of the life forms under consideration). 
Even though UV-BAS or UV-LD can provide information about biological effectiveness of radiation or survival of the population, respectively, there are some drawbacks related to their use. For instance, the values from those extrapolations are taken in general from studies that used low fluence rates\footnote{Defined according to \citet{Braslavsky2007} as: total radiant power, incident from all directions onto a small sphere divided by the cross-sectional area of that sphere. SI unit W m$^{-2}$},in contrast to those expected for more realistic astrophysical scenarios. Another factor, usually not considered, is that the survival of the microorganisms will vary according to their growth stage (e.g.: growth phase) and the physiological conditions of the population (e.g.: starvation). Thus, these kind of extrapolations might not represent accurately the effects of radiation on life considering particular planetary environments and can lead, therefore, to potentially biased results. 
An example of a preliminary laboratory assay have shown an increase in the biomass of microorganisms exposed to UVC fluences comparable to weak flares \citep{Abrevaya2011}. However, these results are only an approximation and cannot be taken as parameter to predict the survival to flares, as the microorganisms were in a physiological condition unfeasible to find in nature and the experiments included the UV-protective effects of the culture medium, among others.
Previously, we proposed how to address the issues derived from the studies above by using an interdisciplinary astrobiological approach that combines astrophysics with microbiology and photobiology in laboratory experiments \citep{Abrevaya2019}. In this Letter, we present the first results and conclusions obtained from this approach, taking as a case study Proxima Centauri $b$ (Proxima $b$), currently, the closest potentially habitable terrestrial-mass planet outside the Solar System, located in the habitable zone of Proxima Centauri (Proxima Cen), an M6V dwarf \citep{Anglada2016}. We applied a radiative transfer modelling to estimate the UVR at the surface of Proxima $b$ during quiescent stellar conditions, taking into account different atmospheric compositions and pressures. Those estimations led us to focus on experiments to determine the biological effects of stellar flares on Proxima $b$, considering unattenuated UVC surface fluxes in a ''worst case scenario''.

This work is part of the EXO-UV programme \citep{Abrevaya2014}, an international interdisciplinary collaboration that seeks to expand the characterization of UVR environments through experimental approaches.

\section{Methods}
\subsection{Estimation of UV surface fluxes during quiescent stellar conditions}
\label{UVRquiescent}
The quiescent stellar spectrum of Proxima Cen was obtained from \citet{Meadows2018} which comprises observational data as well as a model spectrum \citep[PHOENIX 2.0,][]{Husser2013} for spectral regions where no observational data were available. To investigate how much UVR reaches the planetary surface depending on the assumed atmospheric composition, an extended version of the radiative transfer model by \citet{Patel2002} was used, as previously applied for investigations of terrestrial-like atmospheric transmittance on exoplanets orbiting cool white dwarf stars \citep{Fossati2012}. The model calculates the stellar spectrum received at the surface of the planet for a zenith angle of 90$^{\circ}$, i.e. at the equator, in order to provide the case of maximum irradiance.
Very recently, it was predicted that Earth-like atmospheres can not form in the habitable zones of very active stars \citep{Johnstone2019}, and atmospheres containing CO$_2$ are likely stable on water-depleted planets orbiting M dwarfs \citep{Gao2015}. Atmospheres dominated by N$_2$ tend to be unstable under high XUV fluxes and experience efficient atmospheric escape \citep{Tian2008, Lammer2011}. However, sufficient levels of CO$_2$ can protect nitrogen atmospheres from escaping due to infrared cooling \citep{Lichtenegger2010, Johnstone2018}. Based on these studies, a range of realistic atmospheric compositions were considered: a) 100\% CO$_2$, b) 10\% CO$_2$ and 90\% N$_2$, c) 50\% CO$_2$ and 50\%N$_2$ d) 90\% CO$_2$ and 10\% N$_2$, with surface pressures of 100, 200, 500, 1000, 2000, and 5000~mbar.
 
\subsection{Estimation of UV surface fluxes during flares}
\label{UVRflares}
Two stellar flare scenarios for Proxima Cen were considered, a typical flare and a superflare. For the estimation of a typical flare-related UVC flux enhancement for M dwarfs, we used the study of \citet{Welsh2007}, who analysed 52 flares detected by the Galaxy Evolution Explorer in NUV band (175--280~nm), which covers the UVC region. They estimate an average NUV flare energy of $2.5\times10^{30}$~erg for 33 M dwarf flare events, with a range covering $1.2\times10^{28}-1.6\times10^{31}$~erg. Scaling their given flare luminosities to the orbit of Proxima~$b$ (0.0485~AU) yields fluxes in the range of about $5\times10^{-4}$ to 80 ~W~m$^{-2}$ with a mean of ${\sim}7$~W~m$^{-2}$. Our experiments were then conducted with a value of 8.7 ~W~m$^{-2}$, which is comparable to this mean value. Since flares follow a power-law distribution according to their energy, there are more flares with low energies and fewer flares with high energies. Therefore, it is clear that most flares from Proxima Cen will be smaller than our assumed ''typical'' flare value. However, such small flares would provide a negligible addition to the quiescent flux level. Furthermore, we estimated a UVC peak fluence rate of 92 ~W~m$^{-2}$ for the superflare on Proxima based on \citet{Howard2018}. The estimated total energy of this superflare is comparable to the ``great flare'' on AD Leo \citep[$\sim$10$^{34}$erg,][]{Hawley1991, Segura2010}.

\subsection{Microorganisms and irradiation conditions}
\label{microgrow}

To evaluate the potential biological damage of flares on the surface of Proxima $b$, considering the UVC range (as it is the most biologically harmful) in a worst case scenario (unattenauted conditions) we performed experiments with microorganisms from different domains: \textit{Haloferax volcanii}, an archeon inhabiting extreme hypersaline environments, and \textit{Pseudomonas aeruginosa}, an ubiquitous bacterium.
For the irradiation procedure suspensions of the microorganisms were irradiated with UVC tube lamps at fluence rates of 8.7 ~W~m$^{-2}$ (typical flare) or 92 ~W~m$^{-2}$ (superflare). Irradiation times were from 5 to 4500 seconds for the flare, considering the typical distribution for the duration of flares described in \citet{Gershberg2005}, and from 5 to 420 seconds for the superflare, considering the peak of the superflare described in \citet{Howard2018}. The control groups were non-irradiated suspensions.
To test the potential effect of the influence of visible light in the repair capability to DNA damage in the microorganisms, in another experiment, samples of \textit{P. aeruginosa} were exposed for 3 hours to visible light after irradiated at a fluence rate of 92 W m$^{-2}$. The control groups followed the same procedure in the dark. 
Technical details about the irradiation procedure, growth and enumeration of microorganisms, and limit of detection (LOD), can be found in the Appendix. All experiments were performed in triplicate (three independent experiments).

\section{Results and discussion}

\subsection{UV radiation on the planetary surface}

As presented in section \ref{UVRquiescent}, our radiative transfer calculations show the influence of the different atmospheric compositions and pressures on the stellar UVR that reaches the surface of Proxima $b$. Different absorption of the UV (200--380 nm) flux for the different atmospheric scenarios were obtained (see Fig. A, appendix), consistent with previous studies by \citep{Ranjan2017}(see comparison in Fig. B, appendix).

The UVR is dominated by the UVA range (315-400 nm) providing 92\% of the incident radiation. Without atmospheric attenuation the UVA flux is 1.08~W~m$^{-2}$, a value much lower than the UVA flux at the Earth's surface (e.g. 30-50~W~m$^{-2}$ for clear sky). This value would be even lower under the attenuation of the considered atmospheres. Thus, the UVA flux on the surface of Proxima $b$ would not prevent the existence of ''life as we know it''. The effectiveness of UVA to induce lethal damage depends also on mechanisms that in the presence of oxygen produce reactive oxygen species \citep{Kielbassa1997, Oppezzo2011}. According to several studies O-rich atmospheres are not expected to exist in planets orbiting M-type stars \citep{Garcia2017, Johnstone2019}, therefore, this would reduce the damage induced by UVA in this context.

For the UVB (280--315 nm) and UVC (200--280 nm) bands, atmospheric absorption would increase with the CO$_2$ content and with the atmospheric pressure, regardless of the atmospheric compositions (see Fig. A1 and A2, appendix). The flux in these bands would be reduced by several orders of magnitude for CO$_2$ contents of 50\% or more, and for atmospheric pressures higher than 2000 mbar. This is particularly relevant because UVB and UVC radiation are absorbed by nucleic acids, which can undergo chemical modifications and lose their biological functions upon exposure \citep{Kielbassa1997}. Nevertheless, the stellar UVB fluence rate reaching the surface of Proxima $b$ (0.026~W~m$^{-2}$) is lower than that at the Earth's surface which is around 2~W~m$^{-2}$. Therefore, microorganisms would be expected to survive even without the protective effect of an atmosphere. For the particular case of UVC, since the terrestrial atmosphere blocks this wavelength range, the information obtained by experiments involving solar radiation at the Earth's surface can not be used as a reference to evaluate the impact of UVC on life on Proxima $b$. The only data available about biological effects of solar UVC in microorganisms comes from space experiments performed in the low Earth orbit. Several microorganisms exposed to unattenuated extraterrestrial solar UVR spectrum in space experiments, for instance, cells of \textit{Chroococcidiopsis} and spores of \textit{Aspergillus sydowii} and \textit{Aspergillus versicolor} survived being exposed during 1.5 years in the low Earth orbit \citep{Cockell2011, Novikova2015}. In addition, viable spores of \textit{Bacillus subtillis} were recovered after 6 years of exposure to solar UVR in space \citep{Horneck1994}. 

We calculated the solar UVC fluence rate at the top of the atmosphere based on solar spectral irradiance data \citep{Meftah2018} resulting in an UVC fluence rate of 6.27~W~m$^{-2}$, which is around 95 times higher than the UVC flux value estimated at the surface of Proxima $b$ (0.066 W m$^{-2}$, without any atmospheric protection). Therefore, it seems likely that ''life as we know it'' could thrive on the surface of Proxima $b$ under the very low UVC irradiation levels expected during quiescent stellar conditions. 

In addition, if attenuation of UVC by the atmosphere is taken into account as we predicted in our modelling, the combination of some atmospheric compositions and pressures would provide even more protection against the most harmful UV wavelengths for life (Fig. A1 and A2, appendix). In this way, planets with O-rich Earth-like atmospheric compositions are not the only ones offering protection to UVR. This finding expand the results presented in previous studies (see e.g.:\citep{OMalley2019}. 

\subsection{Biological effects of flares}
Given that on the surface of Proxima $b$ the UV fluence rates during quiescence are negligible from the biological point of view, we decided to focus on flare activity and we did not conduct experiments for quiescent levels of UVR. 
Taking into account that UVC is the most deleterious component of UVR, we performed experiments to evaluate the biological damage of UVC fluence rates related to flares on the surface of Proxima $b$, in a worst case scenario (i.e.: no shielding by biological or environmental factors, including no absorption by the atmosphere).

\begin{figure}
   \centering
   \includegraphics[width=7cm]{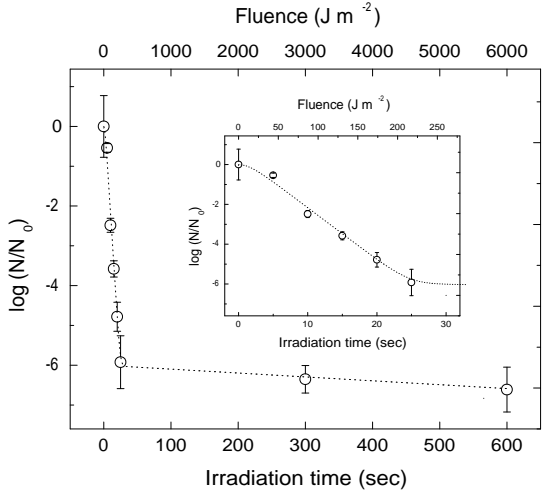}
      \caption{Survival curve for \textit{H. volcanii} exposed to a fluence rate of 8.7 ~W~m$^{-2}$ (typical flare). N$_0$ is the viable count of microorganisms at the beginning of the irradiation, N is the viable count after the irradiated time indicated in the abscissa. The LOD value was -6.76. Error bars indicate three independent experiments. Inset: detail of the curve for the first 15 seconds of irradiation. 
              }
         \label{Fig1}
   \end{figure}

\begin{figure}
   \centering
   \includegraphics[width=7cm]{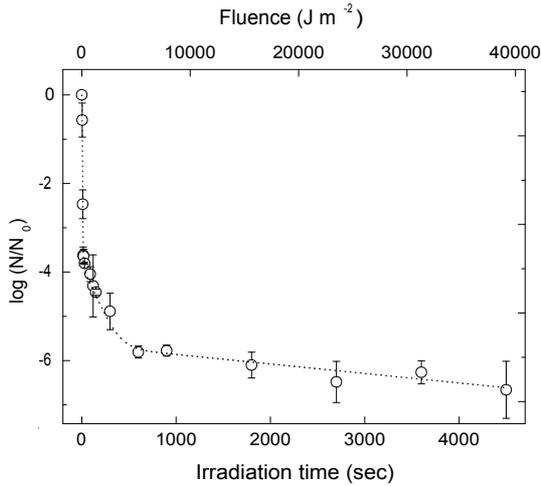}
      \caption{Survival curve for \textit{P. aeruginosa} exposed to a fluence rate of 8.7 ~W~m$^{-2}$ (typical flare). The meaning of N, N$_0$ and error bars is the same as indicated in Fig.2. The LOD value was -7.32.
              }
         \label{Fig2}
   \end{figure}

\begin{figure}
   \centering
   \includegraphics[width=7cm]{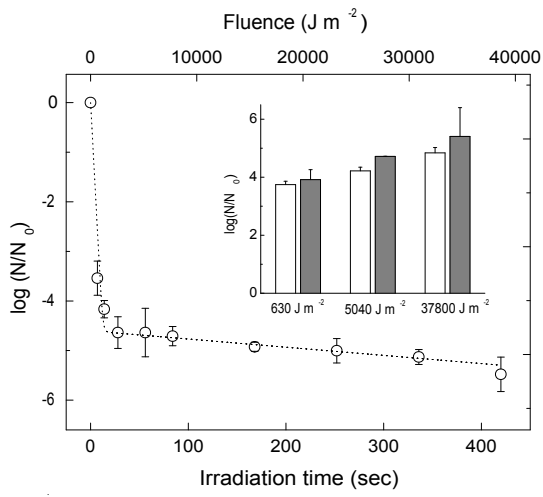}
      \caption{Survival curve for \textit{P. aeruginosa} exposed to a fluence rate of 92 ~W~m$^{-2}$ (superflare). The meaning of N, N$_0$ and error bars is the same as indicated in Fig.2. The LOD value was -7.03. Inset box: Influence of visible light on the survival of irradiated microorganisms. White bars indicate photoreactivating conditions, gray bars correspond to the control group (incubation in dark conditions).
              }
         \label{Fig3}
   \end{figure}

The microorganisms were exposed to fluences comparable to those estimated for the peak of a typical flare (Figs.~\ref{Fig1} and \ref{Fig2}) or a superflare (Fig.~\ref{Fig3}). The maximum irradiation time for the typical flare (Figs.~\ref{Fig1} and \ref{Fig2}) was 4500~s, which corresponds to an energy of ${\sim}2.6\times10^{32}$~erg. For the superflare (Fig.~\ref{Fig3}), the total irradiation time was 420~s, corresponding to a similar energy. The survival curves obtained for the microorganisms exhibit the shape described as typical for the response of a bacterial population to UVC \citep{CoohillandSagripanti2008}. At the beginning of the irradiation, the survival fraction reduced exponentially, as most of the exposed microorganisms lost their viability. 
   
Subsequently, a remarkable reduction in the slopes of the survival curves was observed when irradiation times were prolonged. This change occurred when the survival fraction reached a value of approximately 10$^{-6}$ for \textit{H. volcanii} (after 25~s, i.e. an energy of ${\sim}1.4\times10^{30}$~erg), and the viable counts remained above the LOD up to 600 s (6000 J m$^{-2}$, ${\sim}3.5\times10^{31}$~erg) (Fig.~\ref{Fig1}). In the case of \textit{P. aeruginosa} the slope changed after the same time/energy for the typical flare, and the viable counts remained above the LOD at all considered fluence values (Figs.~\ref{Fig2}). For the superflare (Fig.~\ref{Fig3}), the change of slope occurred at a survival fraction between $10^{-4}$ and $10^{-5}$, i.e. after 28~s (${\sim}1.7\times10^{31}$~erg), and the viable counts remained above the LOD for all fluence values.
This reduction of the slope of the survival curves, usually designated as ''tail'', indicates the presence of minority subpopulations of cells able to tolerate high fluences. The reason for this phenomenon is not fully understood. It has been ascribed to the formation of clumps \citep{CoohillandSagripanti2008}, or alternatively to transitory and reversible phenotypic changes \citep{Pennell2008}, similar to those responsible for persistence in bacteria exposed to bactericidal antibiotics \citep{Lewis2010} and in archaea during starvation conditions or that were treated with various biocidal compounds \citep{MegawandGilmore2017}.

Considering the shape of the survival curves shown in Figs.~\ref{Fig1} to \ref{Fig3}, it seems possible that a small part of a microbial population irradiated during a flare survives. Eventually, survivors could proliferate, reestablishing the microbial population, and therefore the occurrence of a flare would not imply the extinction of the exposed microorganisms. Interestingly, subpopulations able to tolerate high UVC fluences were found even in cultures of a microorganism that can be considered ubiquitous on the Earth surface, like \textit{P. aureginosa}, exposed to a superflare in a ''worst case scenario'' (Fig. \ref{Fig3}). Since survival curves to UVC are frequently biphasic, the effect produced by high fluences, like those expected during flares or superflares, cannot be appraised from the initial slope and from the total fluence imparted, such as previous studies assumed by using the UV-LD. For instance, some of them predicted no survival to superflares taking as parameter for lethality fluences greater than 553 Jm$^{-2}$ \citep{Howard2018,Estrela2018}, which is, according to our experiments, roughly 70 times smaller than the fluence to produce a significant decrease in the viability of the microorganisms on Proxima $b$ during a superflare.
Moreover, previous theoretical studies using data from curves of UV-BAS for microorganisms \citep{Rugheimer2015, OMalley2017, Estrela2018, OMalley2019} or DNA molecules in vitro \citep{Buccino2006, Buccino2007, Rugheimer2015} are missing important information, as they work with UV-BAS usually obtained at low fluences in contrast to those related to flares or superflares. Moreover, when using UV-BAS obtained from DNA molecules there is an additional problem. The results can be unrealistic, because the overall biological effect of UVR will depend not only on the damage to DNA, but also to other cellular structures, as well as on the intrinsic mechanisms of the cell, such as photoprotection and DNA repair. Among these mechanisms, microorganisms could use the energy of the visible light to repair UV-induced DNA damage (i.e.: photoreactivation mechanisms, ~\ref{microgrow}), improving their survival \citep{Sancar2000}. We tested this mechanism in our experiments, for the case of \textit{P. aureginosa} exposed to the superflare, and no significant increase in the survival rate was found when the irradiated bacteria were exposed to visible light in comparison to those incubated in the dark (Fig. \ref{Fig3}, inset box).

\section{Conclusions}
We studied the UV fluxes on the surface of Proxima $b$ to evaluate if this kind of radiation could be a factor to limit the existence of ''life as we know it''.
We demonstrate that other planetary atmospheric compositions than O-rich ones (namely N$_2$ and CO$_2$) can be optically thick to UV, in particular for the most biologically damaging wavelengths (UVC and UVB). Moreover, UV shielding increases as a function of atmospheric pressure. This expands the possible atmospheres to be considered as ''UV-protective'' in habitability studies, beyond O$_2$ - ozone rich atmospheric compositions. Nevertheless, our results show that the UV fluence rates on the surface, considering the star in quiescence, would probably have negligible effects on life even without an atmosphere. Given that, we focused our study on a ''worst case scenario'', considering therefore flare activity in the absence of any sources of UV-shielding, such as an atmosphere. Our experiments demonstrate, for the first time, the potential impact that the UVR from a flare would have on life: when microorganisms receive very high fluences of UVC, such as the amount of radiation expected to reach the surface of Proxima $b$ after a typical flare or a superflare, a fraction of the microbial population is able to survive. How the flare frequency \citep[e.g.][]{Davenport16,Vida2019} would affect the survival of the microbial population remains an open question and will be addressed in a future study. Our present results based on a single flare event show that microorganisms have higher chances to cope with UVR on the surface of Proxima $b$ than estimated in previous studies based on modelling. Beyond the conditions considered in our study, given the potential atmospheric composition and the inherent mechanisms of defense of potential microorganisms against UVR, the probability of life to persist in highly UV irradiated environments could be increased by other biological factors such as clumping and biofilm formation, as well as other environmental factors apart from the atmosphere such as turbidity or the presence of UV-absorbing compounds in a liquid medium or a water column like in an ocean. 
Our study demonstrates that life can cope with very unfavorable conditions that can not be found naturally on our planet. These results can be extended to the UVR surface habitability of other terrestrial planets orbiting M dwarfs. Additionally, they can also be relevant to the study of potential surface biosignatures on other planetary bodies.

\section*{Acknowledgements}
We thank the reviewer, Paul Rimmer, for his valuable comments on this paper. We also thank Dr. H. Lammer for helpful discussions at the beginning of this programme and Dr. S. Ranjan for providing data for model comparisons. X.C.A. acknowledges the Austrian Academy of Sciences (\"OAW) for the Joint Excellence in Science \& Humanities fellowship grant, and partial funding by Proyecto de Unidad Ejecutora IAFE (CONICET) (2016-2021). A.F.F.G. acknowledges funding by Universidad Nacional de Tres de Febrero. M.L. acknowledges the Austrian Science Fund (FWF): P30949-N36.




\bibliographystyle{mnras}
\bibliography{bibfile} 




\appendix

\section{additional information}
Figure A1, A2 and B, and details about methods are available online.

\bsp	
\label{lastpage}
\end{document}